\documentclass[12pt, letterpaper]{article}

\usepackage{amsmath,amssymb}

\usepackage{cite}

\usepackage{fancyhdr}

\usepackage[top=1in, bottom=1in, left=1in, right=1in]{geometry}

\usepackage[dvipdfm, dvips]{graphicx}

\usepackage{hyperref}

\numberwithin{equation}{section}

\newcommand{\dslash}{\not{\hbox{\kern-2pt $\partial$}}}
\newcommand{\bq}{\begin{equation}} 
\newcommand{\eq}{\end{equation}}
\newcommand{\bqa}{\begin{eqnarray}} 
\newcommand{\eqa}{\end{eqnarray}}
\newcommand{\nn}{\nonumber \\}

\newcommand{\JJ}{{\cal J}}
\newcommand{\td}{\tilde}

\begin{document}



\setcounter{page}{0}

\date{}

\lhead{}\chead{}\rhead{\footnotesize{}}\lfoot{}\cfoot{}\rfoot{}

\title{\textbf{TASI Lectures on Emergence of Supersymmetry, Gauge Theory and String in Condensed Matter Systems\vspace{0.4cm}}}

\author{Sung-Sik Lee$^{1,2}$\vspace{0.7cm}\\
{\normalsize{$^1$Department of Physics $\&$ Astronomy, McMaster University,}}\\
{\normalsize{1280 Main St. W., Hamilton ON L8S4M1, Canada}}\vspace{0.2cm}\\
{\normalsize{$^2$Perimeter Institute for Theoretical Physics,}}\\
{\normalsize{31 Caroline St. N., Waterloo ON N2L2Y5, Canada}}}

\maketitle

\thispagestyle{fancy}

\begin{abstract}

\normalsize \noindent 
The lecture note consists of four parts.
In the first part, we review a 
2+1 dimensional lattice model 
which realizes emergent supersymmetry
at a quantum critical point.
The second part is devoted to a phenomenon called fractionalization
where gauge boson and fractionalized particles 
emerge as low energy excitations
as a result of strong interactions between
gauge neutral particles.
In the third part, we discuss about 
stability and low energy effective theory
of a critical spin liquid state
where stringy excitations emerge in a large N limit.
In the last part, we discuss about an attempt to
come up with a prescription to
derive holographic theory for general quantum field theory.
\end{abstract}


\newpage

\tableofcontents

\vspace{1cm}


\section{Introduction}

A variety of phenomena in condensed matter systems
ranging from metal to superconductivity 
can be understood based on simple Hamiltonians,
such as the Hubbard model,
\bqa
H = - \sum_{<i,j>,\sigma} t_{ij} c_{i\sigma}^\dagger c_{j \sigma} + U \sum_i n_{i\uparrow} n_{i \downarrow} - \mu \sum_i n_i.
\eqa
Here $t_{ij}$, $U$ and $\mu$ are hopping integral, on-site Coulomb energy and chemical potential, respectively.
$c_{i\sigma}$ is the annihilation operator of electron with spin $\sigma$ at site $i$, 
$n_{i\sigma} = c_{i\sigma}^\dagger c_{i \sigma}$ and $n_i = n_{i\uparrow} + n_{i\downarrow}$.
While it is easy to write,
it is impossible to 
solve the Hamiltonian of
interacting $10^{23}$ electrons
from the first principle calculation.
The symmetry of the Hamiltonian is low
and there are few kinematical constraints that simplify dynamics.
Because of interactions, 
one can not in general disentangle a subsystem from the rest 
to reduce the many-body problem to one-body problem.
Nonetheless, as one probes the system at low energies,
various dynamical constraints emerge, 
which gives us a chance to understand low energy physics
by focusing on fewer degrees of freedom.
Presumably, the full Hilbert space has a lot of 
local minima, and the system
`flows' to one of the minima in the low energy limit.
Around each minimum, the
geometrical and topological structures of the low energy manifold
determine the nature of low energy degrees of freedom.
Quantum fluctuations of the low energy modes
are governed by low energy effective theories
which are rather insensitive to details
of the microscopic Hamiltonian.
The distinct set of minima and 
the associated low energy effective theories
characterize different phases of matter (universality classes).
One of the goals in condensed matter physics
is to classify different phases of matter
and understand universal properties of them 
using low energy effective theories.

Various quantum phases of matter 
have been identified, 
such as
Landau Fermi liquid, 
band insulator,
superfluid/superconductor,
quantum Hall liquid, etc.
The list is  still growing, and
new ways of characterizing different phases 
are being devised.
However, it is likely that 
there are many new phases yet to be discovered.

There are some interesting aspects we need to appreciate.
First, it is usually very hard to predict
which phase a given microscopic Hamiltonian flows into.
Often, microscopic Hamiltonians do not
give much clue over what emerges in the low energy limit.
Who could have predicted that
the Hubbard model with impurities and defects
has superconducting phase where
electric current can flow without any resistivity?
Second, low energy effective theories in some phases
or quantum critical points in condensed matter systems
are strikingly similar to what (we believe) describes
the very vacuum of our universe\cite{WENBOOK,SUBIR}.
For example, there exists a phase where
the low energy effective theory has very high symmetry 
including supersymmetry even though microscopic Hamiltonian
breaks almost all symmetries except for some discrete
lattice symmetry and internal global symmetry.
In some corner of the landscape, 
gauge theory emerges as a low energy effective theory
of the Hubbard model.
In some phases, there is no well-defined quasiparticle.
Instead, some weakly coupled stringy excitations 
emerge as low energy excitations.
How is it possible that collective fluctuations
of electrons in solids have such striking similarities
to the way the vacuum of our own universe fluctuates ?
Is the universe made of a bunch of non-relativistic `electrons'
at very short distance ?
Perhaps 
it is simply that there 
are not too many good ideas
that are available to nature, and she 
has to recycle same ideas in different systems and different scales 
over and over.
In this lecture, 
we will review a few examples 
of emergent phenomena in condensed matter systems
which are potentially interesting to 
both condensed matter physicists and high
energy physicists.

\section{Emergent supersymmetry}

In spontaneous symmetry breaking,
symmetry of microscopic model
is broken at low energies 
as the ground state spontaneously choose
a particular vacuum among degenerate
vacua connected by the symmetry.
In condensed matter systems,
the opposite situation often arises.
Namely, new symmetry which is absent
in microscopic Hamiltonian can arise
at low energies as the system
dynamically organizes itself 
to show a pattern of fluctuations 
which obey certain symmetry 
in the long distance limit.
Sometimes, gapless excitations (or ground state degeneracy)
whose origins are not obvious from any microscopic symmetry
can be protected by emergent symmetries.

\subsection{Emergence of (bosonic) space-time symmetry}
Consider a rotor model 
in one-dimensional lattice,
\bqa
H_b & = & -t \sum_{i} ( e^{ i ( \theta_i - \theta_{i+1} )} + h.c. ) 
+ \frac{U}{2} \sum_i (n_i- \bar n)^2, 
\eqa
where $\theta_i$ ($n_i$) is phase (number) of bosons at site $i$
which satisfies $[\theta_i,n_j] = i \delta_{ij}$,
and $\bar n$ is the average density.
For integer $\bar n$, 
the long distance physics is captured by the 1+1D XY model
\bqa
S = \kappa \int dx^2 (\partial_\mu \theta)^2,
\eqa 
where $\kappa \sim \sqrt{t/U}$.
Because $\theta$ is compact, 
instanton is allowed 
where the winding number 
\bqa
\nu(t) = \int_{0}^{L} dx \partial_x \theta(t,x)
\eqa
changes by an integer multiple of $2 \pi$.
Here we consider the periodic boundary condition : $e^{i \theta(t,0)} =e^{i \theta(t,L)}$.
Physically, a unit instanton describes quantum tunneling
from a state with momentum $k$ 
to a state with momentum $k+2 \bar n \pi/a$
with $a$ the lattice spacing.
This tunneling is allowed because 
momentum needs to be
conserved only modulo $2\pi/a$ due to the underlying lattice.
\footnote{In the T-dual variable $\partial_\mu \phi = \epsilon_{\mu \nu} \partial_\nu \theta$, 
the non-conservation of the winding number 
translates into the (explicitly) broken translational 
symmetry in the target space $\phi$, 
and the theory is mapped into the sine-Gordon model.}

If $\kappa>>1$,
instantons are dynamically suppressed 
even though no microscopic symmetry prevents them. 
In this case,
the absolute value of momentum is conserved,
not just in modulo $2\pi/a$.
Since the state with momentum $2 \bar n \pi/a$ does not
mix with the state with zero momentum,
states with non-trivial windings arise as well-defined excitations
which becomes gapless excitation 
in the thermodynamic limit.
Note that this gapless excitation is not a Goldstone mode
because the continuous symmetry can not be broken
in 1+1D even at $T=0$.
This gapless mode is a topological excitation
protected by the emergent conservation law.
Within each topological sector, 
we can treat $\theta$ as a non-compact variable
and the low energy theory becomes a free theory.

If $\kappa << 1$, 
the potential energy dominates 
and bosons are localized, exhibiting gap.
In terms of the XY-model,
instantons and anti-instantons proliferate,
and the system can not ignore the 
existence of lattice anymore.

Now let's consider a simple two-dimensional lattice 
model where rotational symmetry emerges,
\bqa
H_b & = & -t \sum_{<i,j>} ( b_i^\dagger b_j + h.c. ) - \mu \sum_i b_i^\dagger b_i,
\eqa
where $t$ is the hopping energy, $\mu$ is chemical potential,
and $<i,j>$ denotes nearest neighbor sites on the square lattice.
The energy spectrum is given by
\bqa
E_k = - 2t ( \cos k_x + \cos k_y ) - \mu.
\eqa
The full spectrum respects only the $90$ degree rotational symmetry.
If the chemical potential is tuned to the bottom of the band,
the energy dispersion of low energy bosons become
\bqa
E_k = t k^2 + O(k^4).
\eqa
In the low energy limit,
higher order terms are irrelevant 
and the full rotational symmetry emerges.

Besides the rotational symmetry, 
the full Lorentz symmetry can emerge.
For example, the low energy excitations of fermions 
at half filling on the honeycomb lattice 
are described by two copies of two-component
Dirac fermions in 2+1 dimensions.
As a result, the full Lorentz symmetry emerges
in the low energy limit although the microscopic
model has only six-fold rotational symmetry.
When there are gapless excitations 
in the presence of Lorentz symmetry, 
usually the full conformal symmetry
is realized.

\subsection{Emergent supersymmetry}

Supersymmetry is a symmetry 
which relates bosons and fermions.
Since bosons and fermions have integer and half
integer spins respectively in relativistic 
systems, supercharges that map boson into fermion
(or vice versa) carry half integer spin
and supersymmetry should be a part of 
space-time symmetry.
Moreover, supersymmetry is the unique 
non-trivial extension of the Poincare symmetry
besides the conformal symmetry.
Given that all bosonic space-time symmetry can emerge
in condensed matter systems, 
one can ask whether supersymmetry can also emerge
\footnote{
There are large literatures on 
supersymmetric quantum mechanics
in condensed matter systems. 
Here we will exclusively focus on
full space-time supersymmetry.}.

In 2D, it is known that 
supersymmetry can emerge 
from the dilute Ising model\cite{FRIEDAN},
\bqa
\beta H = -J \sum_{<i,j>} \sigma_i \sigma_j - \mu \sum_i \sigma_i^2,
\eqa
where $\sigma= \pm 1$ represents a site with spin up or down,
and $\sigma=0$ represents a vacant site.
When $\mu$ is large, almost all sites are filled 
and the usual second order Ising transition occurs
as $J$ is tuned.
As $\mu$ is lowered, the second order transition
terminates at a tricritical point $\mu=\mu_{c}$ 
and the phase transition becomes first order below $\mu_c$.
The tricritical point is described by the 
$\Phi^6$-theory,
\bqa
S = \int d^2x [ (\partial_\mu \Phi)^2 + \lambda_6 \Phi^6 ]
\eqa
where $\Phi \sim < \sigma >$ describes the magnetic order parameter.
Although there is no fermion in this action,
one can construct a fermion field $\psi$
from a string of spins through the Jordan-Wigner transformation.
At the tricritical point, 
the scaling dimensions of $\Phi^2$ and $\psi$ 
differ exactly by $1/2$.
This is not an accident and
these two fields form a multiplet under an emergent supersymmetry.
More generally,
the operators which are even (odd) under the $Z_2$ spin symmetry
form the Neveu-Schwarz (Ramond) algebra.
The dilute Ising model provides deformations
of the underlying superconformal theory 
within $(-1)^F=1$ sector, where $F$ is the fermion number.
For lattice realizations of 
other 2D superconformal field theories, 
see Ref. \cite{Fendley1,Fendley2}.

To realize emergent supersymmetry in higher dimensions, 
it is desired to have an 
interacting theory in the IR limit\cite{SCOTT}.
Otherwise, RG flow of supersymmetry-breaking couplings
would stop below certain energy scale, 
and supersymmetry-breaking terms generically survive
in the IR limit.
It is hard to have free bosons and fermions
which have same velocity 
unless enforced by some exact microscopic symmetry.
In this sense, 2+1D is a good (but not exclusive) 
place to look for an 
emergent supersymmetry.
In 3+1D, it has been suggested that 
the ${\cal N}=1$ super Yang-Mills theory can emerge
from a model of gauge boson and 
chiral fermion in the adjoint representation\cite{KAPLAN}.
\footnote{But the notion of emergent supersymmetry 
is less sharp in this case 
due to the fact that confinement sets in 
below an energy scale
and there is no genuine IR degrees of freedom.
For a given gauge coupling at the lattice scale,
supersymmetry is realized as an approximate 
symmetry between fermionic and bosonic
glue ball spectra.
}
If a subset of supersymmetry
is kept as exact symmetry in lattice models,
a full supersymmetry can emerge in the continuum\cite{CATTERALL}.
Here, we will consider a 2+1D lattice model where supersymmetry 
dynamically emerge at a critical point without any lattice supersymmetry\cite{SUSY}.

\subsubsection{Model}
\begin{figure}[h!]
\centering
        \includegraphics[height=11cm,width=9cm]{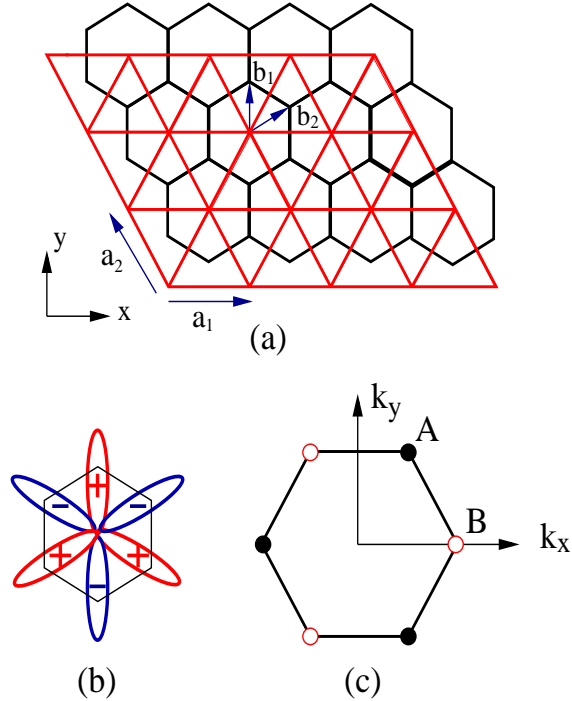}
\caption{
(a) The lattice structure in the real space. 
Fermions are defined
on the honeycomb lattice and 
the bosons, on the dual triangular lattice. 
${\bf a}_1$, ${\bf a}_2$ are the lattice vectors with length $a$, and
${\bf b}_1$, ${\bf b}_2$, two independent vectors
which connect a site on the triangular lattice
to the nearest neighbor sites on the honeycomb lattice.
(b) The phases of a fermion pair in the real space.
(c) The first Brillouin zone in the momentum space.
$A$ and $B$ indicate two inequivalent points with momenta 
${\bf k}_A = \frac{2 \pi}{a} ( \frac{1}{3}, \frac{1}{\sqrt{3}} )$ 
and ${\bf k}_B = \frac{2 \pi}{a} ( \frac{2}{3}, 0 )$ 
where the low energy modes are located.
$\psi_1$, $\phi_2$ are located at ${\bf k}_A$ and
$\psi_2$, $\phi_1$, at ${\bf k}_B$.
}
\label{fig:lattice}
\end{figure}

The Hamiltonian is composed of three parts, 
\bqa
H & = & H_f + H_b + H_{fb},  
\eqa
where
\bqa
H_f & = & -t_f \sum_{<i,j>} ( f_i^\dagger f_j + h.c. ), \nn
H_b & = & t_b \sum_{<I,J>} ( e^{ i ( \theta_I - \theta_J )} + h.c. ) + \frac{U}{2} \sum_I n_I^2, \nn
H_{fb} & = & h_0 \sum_I e^{i \theta_I} (  
f_{I+ {\bf b}_1} f_{I-{\bf b}_1} +     f_{I-{\bf b}_2} f_{I+{\bf b}_2} +  \nn
&& + f_{I-{\bf b}_1+{\bf b}_2} f_{I+{\bf b}_1-{\bf b}_2}     ) + h.c.. 
\eqa
Here $H_f$ describes spinless fermions with nearest neighbor hopping on the honeycomb lattice at half filling;
$H_b$ describes bosons with nearest neighbor hopping and an on-site repulsion on the triangular lattice 
which is dual to the honeycomb lattice; and 
$H_{fb}$ couples the fermions and bosons.
The lattice structure is shown in Fig. \ref{fig:lattice} (a).
$f_i$ is the fermion annihilation operator and
$e^{-i \theta_I}$, the lowering operator of $n_I$
which is conjugate to the angular variable $\theta_I$.
$<i,j>$ and $<I,J>$ denote pairs of nearest neighbor sites 
on the honeycomb and triangular lattices, respectively.
$t_f, t_b  > 0$ are the hopping energies for the fermions and bosons, respectively 
and $U$ is the on-site boson repulsion energy.
${\bf b}_1$ and ${\bf b}_2$ are two independent 
vectors which connect 
a site on the triangular lattice to the neighboring honeycomb lattice sites.
$h_0$ is the pairing interaction strength associated with the process
where two fermions in the f-wave channel around a hexagon 
are paired and become a boson at the center of the hexagon, and vice versa.
In this sense, the boson can be regarded as a Cooper pair 
made of two spinless fermions in the f-wave wavefunction.
This model has a global U(1) symmetry
under which the fields transform as
$f_i \rightarrow f_i e^{i \varphi}$ and 
$e^{-i \theta_I}  \rightarrow e^{-i \theta_I} e^{i 2 \varphi}$.

In the low energy limit,
there are two copies of Dirac fermions
and two complex bosons.
One set of Dirac fermion and complex boson
carries momentum ${\bf k}_A$,
and the other set carries momentum ${\bf k}_B$.
The fermions are massless without any fine tuning,
which is protected by the time reversal symmetry
and the inversion symmetry.
The reason why we have two complex bosons 
instead of one is that the boson kinetic energy is
frustrated.
Because $t_b > 0$, the boson kinetic energy is minimized
when relative phases between neighboring bosons become $\pi$.
However, not all kinetic energy terms can be minimized due to 
the geometrical frustration of the triangular lattice 
where bosons are defined.
The best thing the bosons can do to minimize the kinetic energy
is to form a `spiral wave'  where the phase of bosons
rotates either by $120$ or $-120$ degree around triangles.
There are two such global configurations and they carry the momenta
${\bf k}_A$ and ${\bf k}_B$ respectively.
Each complex boson describes 
condensate of bosons in each of these configurations.
The effective theory for the low energy modes becomes
\bqa
{\cal L} & = & i \sum_{n=1}^2 
{\overline \psi_n} 
\left(
\gamma_0 \partial_\tau +  c_f \sum_{i=1}^2  \gamma_i \partial_i  
\right) \psi_n \nn
 && + \sum_{n=1}^2 
\left[  
 |\partial_\tau \phi_n|^2 + c_b^2 \sum_{i=1}^2 |\partial_i \phi_n|^2
+ m^2 |\phi_n|^2 \right] \nn
&&  + \lambda_1  \sum_{n=1}^2  |\phi_n|^4 
    + \lambda_2  |\phi_1|^2 |\phi_2|^2 \nn
&& + h
 \sum_{n=1}^2 \left( \phi_n^* \psi_n^T \varepsilon \psi_n +   c.c.  \right).
\eqa
A closely related field theory has been studied in the context of high $T_c$ superconductors\cite{Leon98}.
Although there are same number of propagating bosons and fermions\footnote{One complex boson is worthy of 
two components of complex fermions in terms of the number of initial data one has to specify
to solve the equation of motion.},
this Lagrangian contains four supersymmetry breaking terms,
\bqa
m & \neq & 0, \nn
c_b & \neq & c_f, \nn
\lambda_1 & \neq & h^2, \nn
\lambda_2 & \neq & 0.
\eqa
One would naively expect that one has to tune 
at least four parameters to reach the supersymmetric
point.
However, it turns out that one needs to tune only the boson mass
to realize supersymmetry.

\subsubsection{RG flow}
To control the theory, we consider the theory
in $4-\epsilon$ dimensions.
We use the dimensional regularization scheme where
the number of fermion components and the traces of gamma matrices
are fixed.

\begin{figure}[h!]
\centering
        \includegraphics[height=10cm,width=7cm]{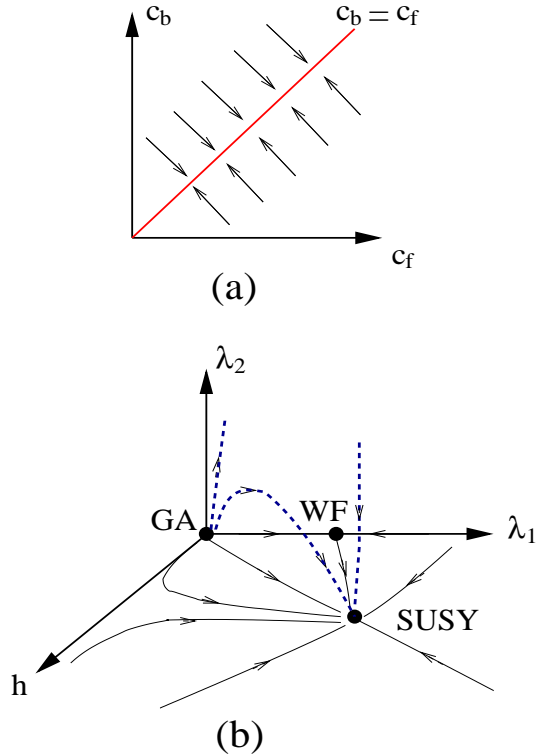}
\caption{
The schematic RG flows of 
(a) the velocities with $h \neq 0$ and
(b) $\lambda_1$, $\lambda_2$ and $h$ in the subspace of $m=0$.
In (b), the solid lines represent the flow in the plane of $(h,\lambda_1)$
and the dashed lines, the flow outside the plane.
}
\label{fig:RG}
\end{figure}

The boson mass is a relevant (supersymmetry-breaking) 
perturbation and 
we tune it (by hand) to zero.
This amounts to tuning one microscopic parameter 
to reach the critical point
which separates the normal phase and 
the bose condensed phase.
The one-loop beta functions for the 
remaining couplings at the critical point 
is given by
\bqa
\frac{d h}{d l} & = & \frac{\epsilon}{2} h - \frac{ 1}{ ( 4 \pi c_f )^2 } 
\left( 2 + \frac{ 16 c_f^3}{ c_b ( c_f + c_b )^2} \right) h^3,  \nn
\frac{d \lambda_1}{d l} & = & \epsilon \lambda_1 - \frac{ 1 }{ ( 4 \pi )^2 } 
\left( 
\frac{ 20 \lambda_1^2 + \lambda_2^2 }{ c_b^2} 
+\frac{ 8 h^2 \lambda_1 }{ c_f^2} 
-\frac{ 16 h^4 }{ c_f^2} 
\right), \nn
\frac{d \lambda_2}{d l} & = & \epsilon \lambda_2 - \frac{ 1 }{ ( 4 \pi )^2 } 
\left( 
\frac{ 4 \lambda_2^2 + 16 \lambda_1 \lambda_2 }{ c_b^2} 
+\frac{ 8 h^2 \lambda_2 }{ c_f^2} 
\right), \nn
\frac{d c_f}{d l} & = & \frac{ 32 h^2 c_f ( c_b - c_f ) }{ 3 ( 4 \pi )^2 c_b ( c_b + c_f )^2}, \nn
\frac{d c_b}{d l} & = & -\frac{ 2 h^2 c_b ( c_b^2 - c_f^2 ) }{  ( 4 \pi c_b c_f )^2 },
\label{beta}
\eqa
where the logarithmic scaling parameter $l$ increases in the infrared.
The RG flow is shown in Fig. \ref{fig:RG}.

The Gaussian (GA) fixed point, $( h^*, \lambda_1^*, \lambda_2^* ) = (0,0,0)$,
the Wilson-Fisher (WF) fixed point, 
$( h^*, \lambda_1^*, \lambda_2^* ) = (0,\frac{( 4 \pi c_b)^2 \epsilon}{20},0)$, 
and the O(4) fixed point,
$( h^*, \lambda_1^*, \lambda_2^* ) = 
(0,\frac{( 4 \pi c_b)^2 \epsilon}{24},\frac{( 4 \pi c_b)^2 \epsilon}{12})$
are all unstable upon turning on the pairing interaction $h$.
If $h$ is nonzero, the boson and fermion velocities begin to
flow as can be seen from the last two equations in Eq. (\ref{beta}).
Because the pairing interaction mixes the velocities of the boson and fermion, the difference of the velocities exponentially flows to zero in the low-energy limit.
The converged velocity in the infrared limit is a non-universal value 
which we will scale to $1$ in the followings.
With a nonzero $h$, the system eventually flows to a stable fixed point,
\bq
( h^*, \lambda_1^*, \lambda_2^* ) = (\sqrt{\frac{(4 \pi)^2 \epsilon}{12}}, \frac{(4 \pi)^2 \epsilon}{12} ,0).
\eq
At this point, the theory becomes invariant under 
the supersymmetry transformation,
\bqa
\delta_\xi \phi_n  =  - {\overline \psi_n} \xi, && \delta_\xi \phi^*_n   =   {\overline \xi} \psi_n \nn
\delta_\xi \psi_n  =  i \slash \hspace{-0.2cm} \partial  \phi_n^* \xi - \frac{h}{2} \phi_n^2 {\cal \varepsilon} {\overline \xi}^T, &&  
\delta_\xi {\overline \psi_n}  =  i {\overline \xi}  \slash \hspace{-0.2cm} \partial \phi_n - \frac{h}{2} \phi_n^{*2} \xi^T {\cal \varepsilon}, \nn
\label{susy}
\eqa
where $\xi$ is a two-component complex spinor.
This theory corresponds to 
the ${\cal N}=2$ (which amounts to four supercharges) Wess-Zumino model
with superpotential,
\bqa
F = \frac{h}{3} ( \Phi_1^3 + \Phi_2^3 ),
\eqa
where $\Phi_1$ and $\Phi_2$ are 
two chiral multiplets. 
Due to the emergent superconformal symmetry, 
the one-loop anomalous scaling dimensions 
for the chiral primary fields $\phi$ and $\psi$,
\bq 
\eta_\phi = \eta_\psi = \epsilon/3
\label{SD}
\eq
are exact.

\begin{figure}[h!]
\centering
        \includegraphics[height=2.5cm,width=8cm]{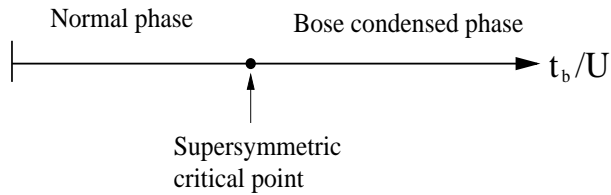}
\caption{
The superconformal theory describes the quantum critical
point separating the normal state and the f-wave 
superconducting (Bose condensed) state of spinless fermions
in the honeycomb lattice.}
\label{fig:phase}
\end{figure}

Although the emergent supersymmetry has been obtained
from a microscopic model which has both fermions and bosons,
one can view bosons as composite particles (Cooper pairs)
which emerge at low energies.
This implies that supersymmetry can emerge
from a microscopic model which contains only fermions.
Therefore, the ${\cal N}=2$ Wess-Zumino theory 
can describe the second order quantum phase transition of
f-wave FFLO (Fulde-Ferrell-Larkin-Ovchinnikov) 
superconducting state of spinless fermions 
on the honeycomb lattice at half filling (see Fig. \ref{fig:phase}).
For an alternative proposal to realize the Wess-Zumino model 
in condensed matter systems, see Ref. \cite{YU}.

\section{Emergent gauge theory}

Fractionalization is a phenomenon 
where a microscopic particle 
in many-body systems 
decay into multiple modes
each of which carry a fractional quantum number
of the original particle.
In contrast to the more familiar phenomena where 
a composite particle 
breaks into its constituent particles at high energies,
fractionalization is a low energy collective phenomenon 
where microscopic particles do not `really' 
break into partons.
Instead, many-body correlations
make it possible for parts of microscopic particles
to emerge as deconfined excitations 
in the low energy limit\cite{ANDERSON}.
When fractionalization occurs,
a gauge field emerges as a collective excitation
that mediates interaction between
fractionalized excitations.
There exist many models which exhibit fractionalization\cite{KITAEV,WEN2003PRL,MOESSNER,WEN2002PRL,WEN2003,MOTRUNICH1,MOTRUNICH2}.
In this lecture, we are going to illustrate
this phenomenon using a simple model\cite{GAUGE}.

\subsection{Model}
Consider a 4-dimensional Euclidean hypercubic lattice (with discretized time).
At each site on the lattice, there 
are boson fields $e^{i \theta^{ab}}$
which carry one flavor index $a$ and one anti-flavor index $b$
with $a,b = 1,2,...,N$.
We impose constraints
$\theta^{ab} = - \theta^{ba}$;
$e^{i \theta^{ab}}$ is anti-particle of 
$e^{i \theta^{ba}}$. 
There are $N(N-1)/2$ independent boson fields per site.
In the following, we will refer to these bosons 
as `mesons'.
The action is
\bqa
S  & = &  -t \sum_{<i,j>} 
\sum_{a,b} 
 \cos \left( \theta^{ab}_{i} - \theta^{ab}_j \right)    \nn
&&  - K \sum_i \sum_{a, b, c} \cos \left( 
\theta^{a b}_i + \theta^{b c}_i + \theta^{c a}_i  \right).
\eqa
Here the first term is the standard Euclidean kinetic energy of bosons
and the second term is a flavor conserving interactions between bosons.
This model can be viewed as a low energy effective theory
of excitons in a multi-band insulator.\footnote{In this context,
an exciton with flavor $a$ and anti-flavor $b$ describes a composite particle 
made of an electron in the $a$-th band and a hole in the $b$-th band
of a multi-band insulator.}
But let's forget about the `true UV theory'
and take this model as our microscopic model
and regard these bosons as fundamental particles.

\subsection{Slave-particle theory}
We are interested in the strong coupling limit ($K >> 1$). 
In this limit, the large potential energy
imposes the dynamical constraints
\bqa
\theta^{a b}_i + \theta^{b c}_i + \theta^{c a}_i =0
\eqa
for every set of $a,b,c$.
The constraints are satisfied on a $(N-1)$-dimensional
manifold in $(S^1)^{N(N-1)/2}$ on each site.
The low energy manifold is parametrized by
\bqa
\theta^{ab}_i = \phi^a_i - \phi^b_i,
\eqa
where the new bosonic fields $\phi^a$ 
carry only one flavor quantum number
contrary to the meson fields.
The new bosons are called slave-particles or partons.
They are `enslaved' to each other 
because these partons
can not escape out of mesons.
Note that there is a local $U(1)$ redundancy 
in this parametrization
and the low energy manifold is
$(S^1)^N/S^1$.

Within this low energy manifold, 
 the potential energy can be dropped
and the kinetic energy becomes
\bqa
S & = &
  -t  \sum_{<i,j>} \left[ \sum_a e^{ i ( \phi^{a}_i - \phi^{a}_j )} \right] 
\left[ \sum_b e^{ - i ( \phi^{b}_i - \phi^{b}_j )} \right]. 
\eqa
Since the kinetic energy is factorized in the flavor space,
one can introduce a collective dynamical field
$\eta \sim  \sum_b e^{ - i ( \phi^{b}_i - \phi^{b}_j )}$
using the Hubbard-Stratonovich transformation.
Then the effective action becomes
\bqa
S & = & 
t \sum_{<i,j>} \left[ 
|\eta_{ij}|^2 
- |\eta_{ij}| \sum_a e^{ i ( \phi^{a}_i - \phi^{a}_j - a_{ij} )} - c.c.
\right],
\eqa
where $a_{ij}$ is the phase of the complex
field $\eta_{ij}$.
Note that the full theory is invariant 
under the U(1) gauge transformation
\bqa
\phi^a_i &\rightarrow & \phi^a_i + \varphi_i, \nn
a_{ij} & \rightarrow & a_{ij} + \varphi_i - \varphi_j.
\eqa
This theory is a compact U(1) lattice gauge theory
coupled with $N$ partons.
The bare gauge coupling is infinite because
the gauge field does not have a bare kinetic energy term.

One may think that
partons are always confined in this theory
because the bare gauge coupling is infinite.
However, we have to be more careful about 
what we mean by confinement.
Since mesons themselves are fundamental particles,
it is indeed impossible 
to literally separate one meson into two partons.
In this sense, a parton is always paired with an anti-parton.
However, this confinement at short distance scale 
does not rule out the possibility
that excitations which carry the same quantum number 
as partons arise as low energy excitations.
If such low energy excitations exist, 
we can say that partons are deconfined
in the long distance limit.

How can we understand that partons are deconfined at low energies ?
To see this, let us integrate out high energy fluctuations of
$\phi^a$ to obtain an effective theory with a cut-off $\Lambda << 1/a$
where $a$ is the lattice spacing.
The fluctuations of partons generate the Maxwell's term (and all
higher order gauge invariant terms) in the low energy effective action,
\bqa
S & = & \int dx^4 \left[
| ( \partial_\mu  - i a_\mu) \Phi_a |^2 + V(\Phi_a) 
+ \frac{1}{g^2} F_{\mu \nu} F^{\mu \nu}  + ... \right],
\eqa
where $\Phi_a = e^{i \phi^a}$.
Due to screening by $N$ charged partons, 
the gauge coupling is renormalized from infinity
down to a finite value $g^2 \sim 1/N$.
In the large $N$ limit, the renormalized 
gauge coupling can be made very  small
and the deconfinement (Coulomb) phase can arise 
as a  stable phase.
Then gapless U(1) gauge boson (emergent photon)
and fractionalized bosons (partons) arise as low energy excitations.

\subsection{World line picture}
Although the above simple argument is plausible, 
the relation between the `slave-partons' which
are confined at short distance scales
and the `liberated-partons' which
are deconfined at long distance scales
is rather obscure in this description.
Moreover, it is not entirely clear how partons can screen 
the gauge field in the first place if they
are always confined within neutral mesons.
In other words, one can question 
how one can integrate out
high energy fluctuations of partons reliably
if they are subject to the infinitely 
strong gauge coupling 
at high energies.

For this reason, it is useful 
to understand fractionalization
purely in terms of 
original mesons without introducing
slave particles. 
Actually one can understand 
fractionalization intuitively
in terms of world lines of mesons.
The world line picture is not a quantitative
description, but 
it is useful in that it provides a physical picture
for fractionalization.
In space-time,
world lines of mesons can be represented
as double lines with two opposite arrows,
where each arrow carries a flavor current.

\begin{figure}
\centering
        \includegraphics[height=8cm,width=8cm]{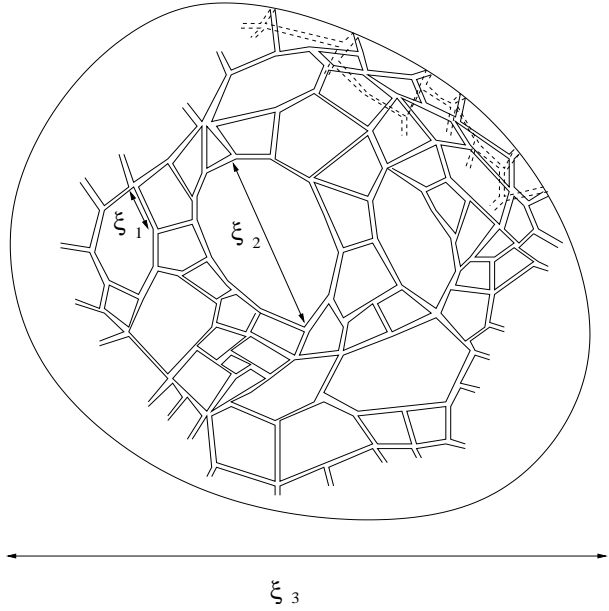}
\caption{
Web of world lines.
}
\label{fig:web}
\end{figure}

In the partition function,
the simplest configuration is 
a loop of meson world line made of 
two overlapped single-line loops 
with current of flavor $a$ moving in one direction and 
$b$ in the other direction.
General configurations consist of a collection of connected webs made of world lines (world line web).
Each world line web defines closed surfaces as depicted in Fig. \ref{fig:web}.
Many such closed surfaces may co-exist and interpenetrate each other.
There are three different length scales in the world line web.
The shortest scale is the length of double line segment ($\xi_1$) 
which corresponds to the life time of a single meson.
As the interaction becomes stronger, mesons decay faster and $\xi_1$ becomes shorter.
The next scale is the typical size of single line loops ($\xi_2$).
This can be much larger than $\xi_1$ 
even in the strong coupling limit.
This is because a single flavor $a$ can switch as many partners
as it wants during its life time before
it meets with its own anti-flavor $\bar a$
and decays into vacuum.\footnote{One's life time can be in principle
much longer than the average lifetime of marriages.}
In this case, we can view single lines as world lines of
emergent long-living excitations.
The largest length scale is the size of the bubble made of the web ($\xi_3$).
By tuning three dimensionless parameters in the theory ($K$, $N$ and $t$),
it is possible to tune three scales, $\xi_1/a$, $\xi_2/a$, $\xi_3/a$,
where $a$ is the lattice spacing.
Here we will focus on the limit $K >>1$ where $\xi_1 \sim a$.\footnote{This
condition is not crucial in what follows. 
In general, one can tune $K$ so that $\xi_1 >> a$.}

In the small $t$ and small $N$ limit 
(which corresponds to the limit of very massive mesons), 
all length scales are of order of the lattice scale.
As $t$ increases, mesons become lighter, and $\xi_2$ increases.
As $N$ increases, configurations with many single line loops
become more important due to entropic contribution of flavors 
associated with each singe line loops.
As a result, the tension of world line web decreases and $\xi_3$ increases.
If $a \sim \xi_1 << \xi_2 << \xi_3$ (which can be achieved for a large $N$ with fixed $tN$) 
the closed surface of the world line web is well-defined in the length scale $\xi_2 << x <<  \xi_3$.
Thus the effective degrees of freedom at this scale
are fluctuating membrane in space-time.
The surface can be regarded as the world sheet of oriented closed string.\footnote{
The world line web is oriented because single lines are oriented.}
In the gauge theory picture the string corresponds to electric flux line 
and the string tension, the gauge coupling.
The gauge group should be U(1) due to the orientedness of surfaces,
and the effective theory becomes a pure U(1) gauge theory.
In this scale, the world line web is very floppy 
and one obtains a weakly coupled pure gauge theory.
The transverse fluctuations of world line web
corresponds to emergent gauge bosons.

For a finite $\xi_3$,
one eventually enters into the region
where $x >> \xi_3$
as one probes longer distance scale.
In this long distance limit, 
there are only small membranes
and the vacuum is essentially empty.
This corresponds to the confinement phase
and $\xi_3$ corresponds the confining scale.

In the opposite limit, if we lower our length scale $x$ 
to the scale comparable to $\xi_2$,
one begins to notice holes on closed surfaces.
One can interpret boundaries of the holes as world lines
of particles.
These particles should be viewed as 
bosons because each contribution adds up 
in the partition function 
without minus signs associated with single line loops.
This is related to the fact that 
$e^{-S}$ and the measure in the partition function
are positive definite in the world line basis
due to the absence of frustration in this model.
The emergent bosons carry unit charge with respect to the 
gauge field.
This is because world line web which corresponds to world sheet 
of one unit of electric flux line can end on boundaries.
Moreover, these bosons carry only one flavor quantum number
due to the fact that an anti-flavor $b$ which always
accompanies a flavor $a$ 
is canceled by the flavor $b$ of nearby mesons.\footnote{Imagine 
that there is a string of mesons, $ab-bc-cd-de-...$.
If you look at this string in the scale larger than the average
spacing between mesons, you will see only flavor $a$ at the end 
of the string.}
This is analogous to the situation 
where one can find a nonzero
charge density on a surface of a dielectric medium made 
of charge neutral molecules.
Therefore in the scale $x \sim \xi_2$, the effective degrees 
of freedom are gauge bosons coupled with $N$ fractionalized
bosons which carry only one flavor quantum number.

In the shortest length scale $x << \xi_2$, 
loops of single lines look very large and
these fractionalized bosons appear to be condensed.
It corresponds to the Higgs phase of the gauge theory.

Therefore, for finite $\xi_2$ and $\xi_3$ 
there are crossovers from the behavior of Higgs phase
to Coulomb phase and eventually to the confinement 
as one probes the system 
at progressively longer distance scales.
The crossover within the confinement phase become
a real phase transition to a deconfinement phase 
as $\xi_3$ is made to diverge
by making $N$ larger with fixed $tN$.
If $\xi_3$ diverges while $\xi_2$ remains finite,
the fractionalized bosons remain gapped but the gauge boson becomes gapless.
This is the Coulomb phase.
\footnote{In 2+1D, Coulomb phase can not be stable with massive partons
due to non-perturbative effects\cite{POLYAKOV77}.
This opens up the possibility that $\xi_2$ and $\xi_3$ diverge
simultaneously without an additional fine tuning
at a critical point\cite{SENTHIL04}.}

Although a parton is always bound with an anti-parton,
it can propagate in space rather freely 
by switching its partners repeatedly in the Coulomb phase.
As a result, partons can be effectively liberated
and arise 
as well-defined excitations in the Coulomb phase.

One may ask why one gets a weakly coupled gauge theory 
rather than some string theory.
One possible answer is that world line web is very dense 
in space-time in this case.
They constantly join and split, 
and their contact interactions are also important.
It is hard to imagine to have a well defined string
in this strongly coupled soup of strings.
This can be viewed as a Lagrangian (or space-time) 
picture for the string-net condensation\cite{LEVIN}.

One can study different types of flavor preserving interactions.
In particular, one can obtain fractionalized fermions\cite{LEVIN,GAUGE} 
if one introduces a quartic interaction for mesons,
\bqa
K_4 \sum_{a,b,c,d} \cos( \theta^{ab}+\theta^{bc}+\theta^{cd}+\theta^{da})
\eqa
instead of the cubic interaction.
If $K_4 > 0$, the interaction is frustrated and
not all terms in the potential energy can be minimized
simultaneously.
This frustration results in a larger
low energy manifold and wavefunctions within
the low energy manifold acquires non-trivial
sign structure, namely wavefunctions describing
low energy excitations are no longer
positive definite.
This non-trivial sign structure is responsible 
for the fermionic statistics
of emergent excitations.
The emergent fermion can be also understood 
from the world line picture where
single line loops are endowed with minus signs
in the partition function due to the frustrated interaction.
Then one should interpret single line loops
as world lines of fermions.

\section{Critical spin liquid with Fermi surface}
\subsection{From spin model to gauge theory}

\subsubsection{Slave-particle approach to spin-liquid states}
Let us consider a system where spins
are antiferromagnetically coupled
in a two-dimensional lattice,
\bqa
H = J \sum_{<i,j>} \vec S_i \cdot \vec S_j + ...
\eqa
Here we focus on models with spin $1/2$.
The antiferromagnetic coupling $J>0$ favors 
neighboring spins that align in anti-parallel
directions.
The dots denote higher order spin interactions
whose specific forms are not important for the following discussion.
The spin model can be viewed as a low energy effective theory
of the Hubbard model at half-filled insulating phase
where charges can not conduct due to a large Coulomb repulsion.

At sufficiently low temperatures $T < J$, 
spins usually order,
spontaneously breaking the SU(2) spin 
rotational symmetry.\footnote{We can treat the SU(2) symmetry as an internal symmetry
in the absence of the spin-orbit coupling which is typically small.}
However, magnetic ordering can be avoided even at zero 
temperature if quantum fluctuations are strong enough.
Two important sources for strong quantum fluctuations 
are proximity to metallic state and geometrical frustration.
In systems which have small charge gap, 
the ... terms induced by charge fluctuations
cause quantum fluctuations of spins
(such as permutations of spins around plaquettes)
which tend to disrupt magnetic ordering.
Quantum fluctuations are also enhanced by
geometrical frustrations.
Geometrical frustrations arise when 
no spin configuration can
minimize all interaction terms simultaneously
in the classical Hamiltonian.
For example, antiferromagnetic couplings 
on the triangular lattice
can not be simultaneously minimized.
Although there is no magnetic long-range order
in disordered ground states,
spins remain highly correlated
and we refer to such correlated non-magnetic
state as spin liquid\cite{SL_REVIEW} as opposed to spin gas.
More importantly, many-body wavefunctions of spin liquids 
exhibit non-local entanglement\cite{Hastings2004}
which can be captured only through
a new quantum mechanical notion of order\cite{Wen_SL}
which is beyond the scheme of conventional order parameter.

One way of understanding spin liquid states 
is to use slave-particle approach.
For example, one decomposes a spin operator into fermion bilinear
\bqa
\vec S_i = \sum_{\alpha \beta} f_{i \alpha}^\dagger \vec \sigma_{\alpha \beta} f_{i \beta}.
\eqa
Here $f_{i \alpha}$ is a fermionic field
which does not carry electromagnetic charge 
but carry only spin $1/2$.
For this reason, this particle is called `spinon'
This decomposition has the U(1) phase redundancy.
As a result, the theory for spinon has to be
in the form of gauge theory.
A compact U(1) gauge theory coupled with 
spinons can be derived following a similar 
step described in the previous lecture.
Although the bare gauge coupling is infinite,
the coupling is renormalized
to a finite value in the low energy limit.
If deconfinement phase is stabilized, 
spinons and the gauge field arise
as low energy degrees of freedom.

Since there is spin $1/2$ per each site,
there is one spinon per site.
Although there is no bare kinetic term
for spinons, they can
propagate in space through 
the exchange interaction :
simultaneous flips of neighboring spins can be
viewed as two spinons exchanging their positions.
As a result, spinons form a band,
which then determines
the low energy spectrum. 
In non-bipartite lattice 
(a lattice which can not be divided into A and B sublattices,
such as the triangular lattice),
the fermions generically form a Fermi surface at half filling.
In this case, the low energy effective theory
has a Fermi surface of spinons 
coupled with the emergent U(1) gauge field\cite{MOTRUNICH,LEE_U1},
\bqa
S & = & \int d^3 x \Bigl[
\Psi_j^* (  \partial_0 - i a_0 - \mu_F ) \Psi_j \nn
&& + \frac{1}{2m} \Psi_j^* ( -i {\bf \nabla } -  {\bf a})^2 \Psi_j 
 + \frac{1}{4g^2} f_{\mu \nu} f_{\mu \nu}
\Bigr].
\eqa
Here $\Psi_j$ is the fermion field with two flavors, $j=1,2$ (spin up and down)
and $a_\mu = (a_0, {\bf a})$ is the U(1) gauge field with $\mu=0,1,2$.
$\mu_F$ is the chemical potential and
$g$, the gauge coupling.
This is nothing but the three dimensional quantum electrodynamics
(QED3) with a nonzero chemical potential.

The zeroth order question one has to ask
is the stability of the deconfinement phase.
Because of the presence of underlying lattice, 
the U(1) gauge field is compact.
This allows for instanton (or monopole) 
which describes an event localized in time 
where the flux of the gauge field changes by $2 \pi$.
It is known that if there is no gapless spinon,
instantons always proliferate in space-time, 
resulting in confinement\cite{POLYAKOV77}.\footnote{An 
exception is when the time-reversal symmetry is broken.
In this case, the Chern-Simons term suppresses instanton,
stabilizing chiral spin liquid state\cite{WEN1989}.
Another route of stabilizing deconfinement phase is to
break the U(1) gauge group into a discrete gauge group,
such as $Z_2$\cite{ReadSachdev91,SENTHIL2000}.
}
In this case, spin liquid states are not stable 
and spinons are confined.
In the presence of gapless spinons, it is possible that
the gauge field is screened and instanton becomes irrelevant
in the low energy limit\cite{IOFFE}.
If this happens, fractionalized phase is stable 
and spinons arise as low energy excitations.

If there are a large number of gapless spinons 
which have the relativistic dispersion, 
instanton acquires a scaling dimension proportional
to the number of flavors $N$\cite{Murthy,Borokhov,Max08,HERMELE}
and the fractionalized phase is stable.
In the presence of spinon Fermi surface,
it turns out that the deconfinement phase can be stable
for any nonzero fermion flavor.

\subsubsection{Stability of deconfinement phase in the presence of Fermi surface}
To show that instanton is irrelevant in the presence of
Fermi surface, we need the following
four ingredients\cite{DECON}.

First,
one can describe low energy particle-hole excitations near the Fermi surface
in terms of an infinite copy of 1+1 dimensional fermions parametrized
by the angle around the Fermi surface\cite{SHANKAR}.
In this angular representation, 
instanton becomes a twist operator 
which twists the boundary condition 
of the 1+1D fermions by $\pi$.
In the presence of an instanton,
the fermions become anti-periodic 
around the origin in the 1+1D space-time.
This can be understood in the following way.
At each point on the Fermi surface,
the Fermi velocity is perpendicular to the Fermi surface.
A fermion (or a wave packet made of states near a point on the Fermi surface) 
with a given Fermi velocity explores
a 1+1 dimensional subspace in real space which is 
perpendicular to the x-y plane in 2+1D.
Since instanton is a source of $2\pi$ flux 
localized in space-time,
the total flux of $\pi$
penetrates through this plane.
(Here we are assuming that there is 
the rotational symmetry.
But this argument can be generalized to cases without the symmetry.)
Therefore, a fermion moving in the plane encloses
a flux close to $\pi$ as it 
is transported around the origin 
at a large distance.
As a result, the boundary condition of low energy fermion
at each angle is twisted by $\pi$.

Second, the angle $\theta$ around the Fermi surface 
has a positive scaling dimension.
As a result,  the angle is decompactified
and runs from $-\infty$ to $\infty$ in the low energy limit.
In the low energy limit, the gauge field becomes more and more ineffective in scattering fermions from one momentum to another momentum along tangential directions to the Fermi surface.
This is  because the momentum of the gauge field becomes 
much smaller than the Fermi momentum.
This amounts to saying that effective angular separation between 
two fixed points on the Fermi surface grows
in the low energy limit.

Third, the theory is local in the space of the decompactified angle.
This is due to the curvature of the Fermi surface.
In the low energy limit, only those fermions very close to the Fermi surface
can be excited.
Since gauge boson scatters one fermion near the Fermi surface 
to another point near the Fermi surface,
the momentum of gauge boson is almost tangential to the Fermi surface
where the fermions are located.
Therefore, fermions with a particular angle are coupled only
with those gauge field whose momentum is tangential to the
Fermi surface at the angle in the low energy limit.
This means that fermions with a finite angular separation
(except for those which are at the exact
 opposite sides of the Fermi surface)
becomes essentially decoupled in the low energy limit
because they are coupled only with those gauge bosons 
whose momenta are separated in the momentum space.

\begin{figure}
\centering
        \includegraphics[height=6cm,width=8cm]{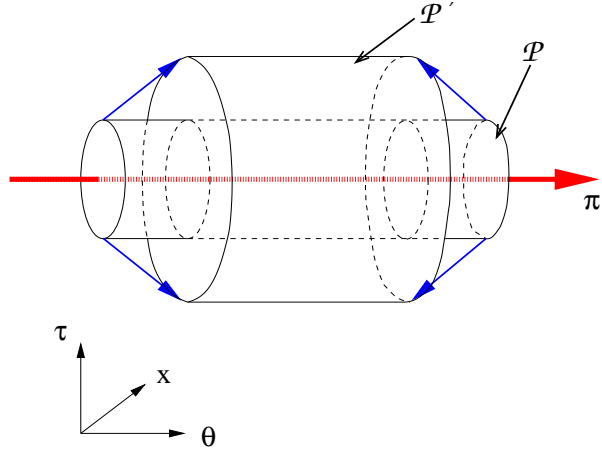} 
\caption{
`Time'-evolution of a quantum state defined on the surface of 
a pipe extended along the angular direction, where
a $\pi$-vortex is pierced through the pipe.
Under the time-evolution, a point on the surface ${\cal P}$ 
is mapped to a point on the surface ${\cal P^{'}}$. 
 }
\label{fig:scale}
\end{figure}

Finally, 
instanton become a $\pi$ vortex which is extended along the
non-compact $\theta$ direction 
in the theory written in the space of $\theta$, $\tau$ and $x$,
where $x$ is a real space coordinate associated 
with the radial momentum 
at each point on the Fermi surface.
In analogy with the state-operator correspondence in relativistic CFT,
the vortex  defines a quantum state on the surface of a pipe ${\cal P}$ 
which is extended in the $\theta$ direction 
in the space of $\tau$, $x$ and $\theta$ 
as is shown in Fig. \ref{fig:scale}.
The scaling dimension of instanton 
corresponds to the `energy' of this quantum state 
associated with the radial time evolution. 

Because of the locality along the decompactified angular direction, 
the extended $\pi$-vortex 
should have an infinite `energy'.
This implies that the scaling dimension of instanton diverges 
and instanton remains strongly irrelevant.
As a result, the deconfinement phase is stable.

\subsection{Low energy effective theory}

Having established that deconfinement phase is stable,
we can treat the theory as a non-compact U(1) gauge theory.
In this lecture, we are going to analyze the low energy effective theory
of Fermi surface coupled with the U(1) gauge field in 2+1D\cite{PLEE89,Halperin,PLEE92,POLCHINSKY,KIM94,NAYAK,ALTSHULER,MotrunichFisher,FS,MAX}.
Although we motivated this theory in the context of spin liquid,
the same theory often arises in other systems.
More importantly, the theory represents 
a class of typical non-Fermi liquid states 
which arise as a result of coupling between Fermi surface 
and gapless bosons.
Not surprisingly, many features of this theory
are shared by other non-Fermi liquid states in 2+1D\cite{O1,O2,O3,O4,O5,O6,O7,O8,O9}.

In this system,
there is no tunable parameter
other than the number of fermion (vector) flavors $N$.
Given that the physically relevant theory ($N=2$)
is a strongly coupled theory,
it is natural to consider the theory 
with a large number of fermion flavors $N$.
Naively one would expect that the theory becomes 
classical in the large $N$ limit.
However, this intuition based on relativistic
field theories is incorrect in the presence of Fermi surface.
Unlike those theories where gapless excitations
are located only at discrete set of points in the momentum space,
Fermi surface has an extended manifold of gapless points.
The abundant low energy excitations subject
to the strong IR quantum fluctuations in 2+1D 
make the theory quite nontrivial 
even in the large N limit\cite{FS}.

In the low energy limit, fermions 
whose velocities are not parallel 
or anti-parallel to each other 
are essentially decoupled.
As a result, 
it is sufficient to consider local patches 
of Fermi surface in the momentum space.
The minimal model in parity symmetric systems
is the theory which includes an open patch 
of Fermi surface and one more patch whose 
Fermi velocity is opposite to that of the former.

\begin{figure}
\centering
        \includegraphics[height=4cm,width=5cm]{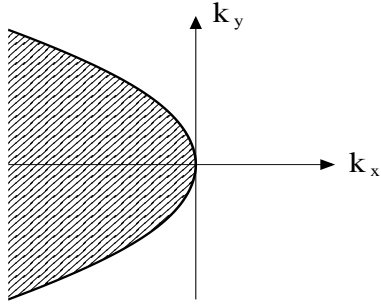} 
\caption{
The parabolic Fermi surface of the model in Eq. (\ref{a1}).
The shaded region includes negative energy states.
}
\label{fig:FS}
\end{figure}

Here, we will focus on one patch in the momentum space.
The one-patch theory is already quite non-trivial
and the full structure is yet to be understood.
The two-patch theory has yet another level of complications\cite{MAX}
over which we have much less theoretical control.
The action in the one-patch theory is
\bqa
{\cal L} & = & \sum_j 
\psi^*_{j} ( \partial_\tau - i v_x \partial_x - v_y \partial_y^2 ) \psi_{j} \nn
&& + \frac{e}{\sqrt{N}} \sum_j  a \psi^*_{j} \psi_{j} \nn
&& + a \left[  -\partial_\tau^2 - \partial_x^2 - \partial_y^2 \right] a,
\label{a1}
\eqa
where $\psi_{j}$ represent fermions with flavor $j=1,2,...,N$.
We have chosen  our ${\bf k}=0$ to be the point on the Fermi surface
where the Fermi velocity is parallel to the $x$-direction.
$v_x$ is the Fermi velocity and $v_y \sim \frac{1}{m}$ determines 
the curvature of the Fermi surface.
The Fermi surface is on $v_x k_x +  v_y k_y^2 = 0$ as is shown in Fig. \ref{fig:FS}.
This is a `chiral Fermi surface' where 
the x-component of Fermi velocity is always positive.
This chirality is what makes the one-patch theory
more tractable compared to the two-patch theory.
$a$ is the transverse component of an emergent U(1) gauge boson
in the Coulomb gauge $\nabla \cdot {\bf a} = 0$.
We ignore the temporal component of the gauge field which
becomes massive due to screening.
The transverse gauge field remains gapless.
$e$ is the coupling between fermions and the critical boson.

Quantum fluctuations of gapless modes generate 
singular self energies.
The one-loop quantum effective action becomes
\bqa
\Gamma & = &  \sum_{j} \int dk  
\Bigl[ i \frac{c}{N} ~\mbox{sgn}(k_0) |k_0|^{2/3} 
+  i k_0  +  v_x k_x + v_y k_y^2 \Bigr] \psi_{j}^*(k) \psi_{j}(k)  \nn
&& + \int  dk 
 \left[ \gamma \frac{|k_0|}{|k_y|} +  k_0^2 + k_x^2 +  k_y^2 \right] a^*(k) a(k) \nn
&& + \frac{e}{\sqrt{N}} \sum_{j}  \int dk dq ~~  a(q) \psi^*_{j}(k+q) \psi_{j}(k),
\label{ga}
\eqa 
where $c$ and $\gamma$ are constants of the order of $1$.
In the low energy limit, 
the leading terms of the quantum effective action are invariant under the scale transformation,
\bqa
k_0 & = & b^{-1} k_0^{'}, \nn
k_x & = & b^{-2/3} k_x^{'}, \nn
k_y & = & b^{-1/3} k_y^{'}, \nn
\psi_{a}(b^{-1} k_0^{'}, b^{-2/3} k_x^{'}, b^{-1/3} k_y^{'}) & = & b^{4/3} 
\psi_{a}^{'}( k_0^{'},  k_x^{'}, k_y^{'}).
\label{scale3}
\eqa
Dropping terms that are irrelevant under this scaling,
we write the minimal action as
\bqa
{\cal L} & = & \sum_j \psi^*_{j} 
( \eta \partial_\tau - i v_x \partial_x - v_y \partial_y^2 ) \psi_{j} \nn
&& + \frac{e}{\sqrt{N}} \sum_j  a \psi^*_{j} \psi_{j} 
 + a ( - \partial_y^2 ) a.
\label{a3}
\eqa
Note that the local time derivative term in the fermion action
is also irrelevant,
and $\eta$ flows to zero in the low energy limit.
But we can not drop this term from the beginning.
Otherwise, the theory becomes completely local in time 
and there is no propagating mode.
The role of this irrelevant $\eta$-term 
is to generate 
a non-trivial frequency dependent dynamics
by maintaining the minimal causal structure of the theory
before it dies off in the low energy limit.
In other words, the $\eta$-term itself is irrelevant
but it is crucial
to generate singular self energies.
But once we include the frequency dependent self energies,
we can drop the $\eta$-term as far as 
we remember that the non-local self energies 
have been dynamically generated from the local Lagrangian.

The minimal action (\ref{a3}) has four marginal terms.
On the other hand, there are five parameters that set 
the scales of energy-momentum
and the fields.
Out of the five parameters, only four of them can modify 
the coefficients of the marginal terms
because the marginal terms remain invariant under 
the transformation (\ref{scale3}).
Using the remaining four parameters, 
one can always rescale 
the coefficients of the marginal terms
to arbitrary values.
Therefore, there is no dimensionless parameter in 
this theory except for the fermion flavor $N$.
In particular, the gauge coupling $e$ can be always
scaled away.
This implies that the theory with fixed $N$ 
flows to a unique fixed point 
rather than a line of fixed point which
has exact marginal deformation.
In the following, we will set $v_x = v_y=e=1$.

\subsubsection{Failure of a perturbative $1/N$ expansion}

In the naive $N$ counting,
a vertex contributes $N^{-1/2}$ 
and a fermion loop contributes $N^1$.
In this counting, 
only the fermion RPA diagram 
is of the order of $1$,
and all other diagrams are of higher order in $1/N$.
In the leading order, the propagators become
\bqa
g_0(k) & = & \frac{1}{ i \eta k_0 +  k_x + k_y^2 } , \nn
D(k) & = & \frac{1}{ \gamma \frac{|k_0|}{|k_y|} +  k_y^2 }.
\label{propagators}
\eqa
One can attempt to compute the full quantum effective action
by including $1/N$ corrections perturbatively.
However, it turns out vertex functions 
which connect fermions on the Fermi surface
generically have strong IR singularity,
which is cured only by loop corrections.
Therefore it is crucial to include
the fermion self energy
in the dressed fermion propagator,
\bqa
g(k) = \frac{1}{ 
i \eta k_0 + i \frac{c}{N} ~\mbox{sgn}(k_0) |k_0|^{2/3} 
+ k_x + k_y^2 }.
\eqa
Although the self energy has an additional factor of
$1/N$ compared to the bare frequency dependent term,
the self energy dominates at sufficiently low energy
$k_0 < 1/(\eta N)^3$ for any fixed $N$.
Here it is important to take the low energy limit first
before taking large $N$ limit.
This is the correct order of limits to probe
low energy physics 
in any physically relevant systems
with finite $N$.

The removal of IR singularity does 
not come without price.
In the presence of the self energy,
the IR singularity is cut-off 
at a scale proportional to $1/N$.
As a result 
the IR singularity
is traded with a finite piece 
which has an additional positive power of $N$.
\begin{figure}[h]
\centering
        \includegraphics[height=5cm,width=7cm]{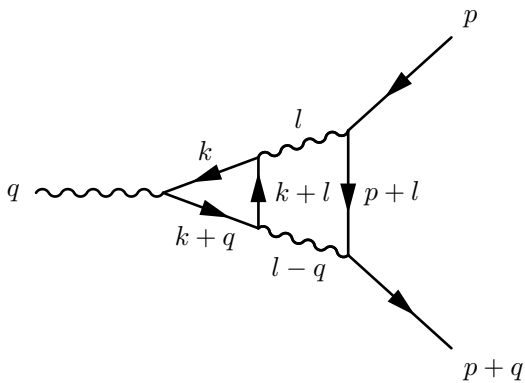} 
\caption{A two-loop vertex correction.}
\label{fig:vertex1}
\end{figure}
For example, the two-loop vertex function  
( Fig. \ref{fig:vertex1} )
computed using the bare propagator
becomes
\bqa
\Gamma^{bare}(p=0,q) = -\frac{N^{-3/2}}{\eta q_0^{1/3}} f_1( q_y/q_0^{1/3} )
\label{eq:vertex2}
\eqa
when both both $p=0$ and $p+q$ are on the Fermi surface.
Here $f_1(t)$ is a non-singular universal function 
which is independent of $N$ and $\eta$.
Note that the vertex function diverges as $q$ goes to zero with fixed $q_y/q_0^{1/3}$.
With the dressed propagator,
the IR singularity disappears
but there exists an enhancement factor of $N$,
\bqa
\Gamma^{dressed}(0,q) = - N^{-1/2} f_2(q_y/q_0^{1/3} ),
\label{eq:vertex3}
\eqa 
where $f_2(t)$ is a non-singular universal function independent of $N$.
Note that this two-loop vertex correction has 
the same order as the bare vertex,
which signals a breakdown of perturbative $1/N$ expansion.
Similar enhancement factors arise in other diagrams as well.

\subsubsection{Genus Expansion}

There exists a simple way of understanding
this enhancement factor systematically.
The reason why the enhancement factor arises
is that the fermion propagator is not always order of $1$.
For generic momentum away from the Fermi surface,
the kinetic energy dominates and
the propagator is order of $1$.
On the other hand, when fermions are right on the 
Fermi surface, the kinetic energy vanishes
and there are only frequency dependent terms.
At sufficiently low frequencies, 
the non-local self energy dominates
and the propagator is enhanced to the order of $N$.
In relativistic field theories,
there are only discrete set of points 
in the momentum space where the kinetic 
energy vanishes.
In contrast,
there is one dimensional manifold of gapless points
in the present case with Fermi surface.
Whenever fermions hit the Fermi surface 
(there are many ways to do that), 
the fermion propagator gets enhanced to order of $N$.
This is the basic reason why the naive $N$ counting breaks 
down in the presence of Fermi surface.

For a given diagram, say $L$-loop diagram,
there are $2L$ integrations of internal momenta $k_x$ and $k_y$.\footnote{The frequency integrals
do not play an important role as far as $N$ counting is concerned.}
In the $2L$-dimensional space of internal momenta,
in general 
there is a $m$-dimensional sub-manifold
on which all internal fermions stay
right on the Fermi surface
as long as external fermions are on the Fermi surface.
We refer to this manifold as
`singular manifold'.
If we focus on the momentum integration near a generic point
on the singular manifold,
it generically looks like
\bqa
I & \sim & 
\int dq_1 dq_2 ... dq_{m} \int dk_1 dk_2 ... dk_{2L-m} 
\Pi_{i=1}^{I_f}
\left[ \frac{1}{ \alpha^i_j k_j + i \frac{1}{N} f( \omega_i) } \right].
\eqa
Here $q_1, q_2, ..., q_m$ are deviation of momenta 
from the point on the singualr manifold 
along the directions tangential to the singular manifold.
$k_1, k_2, ..., k_{2L-m}$ are momentum components
which are perpendicular to the singular manifold.
$I_f$ is the number of fermion propagators.
The key point is that the fermion propagators
depend only on $k$'s but not on $q$'s
because kinetic energy of fermions stay zero
within the singular manifold.
The tangential momenta $q$'s parametrize 
exact zero modes of Fermi surface deformations
where fermions on the Fermi surface
slide along the Fermi surface.
If $N$ is strictly infinite
and we drop the frequency dependent term in the propagators,
the fermion propagators become singular whenever 
fermions are on the Fermi surface.
The $m$ integrations along the tangential direction
of the singular manifold do not help to remove 
the IR singularity.
Only $2L - m$ integrations of $k$ momenta
lower the degree of IR singularity.
After the integration over the all $2L$ momentum,
the remaining IR singularity is order of $I_f - (2L-m)$.
For a finite $N$, this IR singularity
is cut-off at a momentum proportional to $1/N$.
This means the resulting diagram has
an additional factor of $N^{I_f - (2L-m)}$ 
as compared to the naive $N$ counting.

\begin{figure}
\centering
        \includegraphics[height=3.5cm,width=8cm]{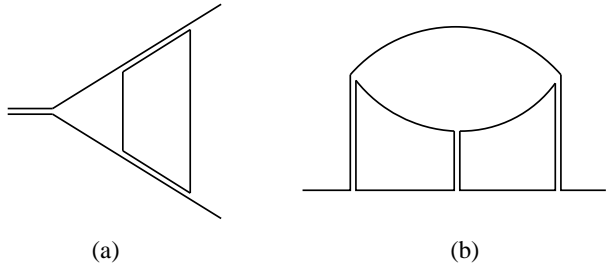} 
\caption{
The double line representations of 
(a) the 2-loop vertex correction and 
(b) the 3-loop fermion self energy.
Double lines represent propagators of the boson, 
and the single lines are the propagators of the fermion.
The number of single line loops (one in (a) and two in (b)) 
represents the dimension of the singular manifold (see the text) 
on which all fermions remain on the Fermi surface
in the space of internal momenta. 
}
\label{fig:vertex_fse_double}
\end{figure}

What determines the dimension of the singular manifold
within which fermions always remain on the Fermi surface?
It turns out that the dimension of 
the singular manifold is given by
the number of closed loops 
when one draws boson propagators
using double lines 
and fermion propagators 
using single lines\cite{SHANKAR,TSAI}.
First, we restrict momenta of all fermions to be on the Fermi surface.
A momentum ${\bf k}_\theta$ of fermion on the Fermi surface
is represented by an one-dimensional parameter $\theta$.
Then, a momentum of the boson $\textbf{q}$
is decomposed into two momenta on the Fermi surface
as ${\bf q}={\bf k}_\theta-\textbf{k}_{\theta^{'}}$,
where both $\textbf{k}_\theta$ and $\textbf{k}_{\theta^{'}}$ 
are on the Fermi surface.
This decomposition is unique
because there is only one way of choosing 
such $\textbf{k}_\theta$ and $\textbf{k}_{\theta^{'}}$.
As far as momentum conservation is concerned,
one can view the boson of momentum $\textbf{q}$
as a composite particle made of 
a fermion of momentum $\textbf{k}_\theta$
and a hole of momentum $\textbf{k}_{\theta^{'}}$.
For example, the two-loop vertex correction 
and the three-loop fermion self energy correction 
can be drawn as Fig. \ref{fig:vertex_fse_double}
in this double line representation.
In this representation, each single line
represents a momentum on the Fermi surface.
Momenta in the single lines that are connected 
to the external lines should be uniquely fixed
in order for all fermions to stay on the Fermi surface.
On the other hand, momenta on 
the single lines that form closed loops
by themselves are unfixed.
In other words, all fermions can stay on the Fermi surface
no matter what the value of the unfixed momentum component
that runs through the closed loop is.
Since there is one closed loop in Fig. \ref{fig:vertex_fse_double} (a),
the dimension of the singular manifold is $1$
and the enhancement factor becomes $N^{4-(4-1)}=N$
for the two-loop vertex correction.
Likewise, the enhancement factor for the three-loop fermion self energy
becomes $N^{5-(6-2)}=N$ which makes the three-loop fermion self
energy to have the same power $1/N$ as the one-loop correction.

The enhancement factor is
a direct consequence of the presence of 
infinitely many soft modes associated with
deformations of Fermi surface.
The extended Fermi surface makes it possible
for virtual particle-hole excitations
to maneuver on the Fermi surface
without costing much energy.
As a result, quantum fluctuations becomes 
strong when external momenta are arranged 
in such a way that there are sufficiently 
many channels for the virtual particle-hole 
excitations to remain on the Fermi surface.
This makes higher order processes to be 
important even in the large $N$ limit.
We note that this effect is absent in relativistic quantum field theories 
where gapless modes exist only at discrete points 
in the momentum space.

\begin{figure}[h]
\centering
        \includegraphics[height=4.5cm,width=5cm]{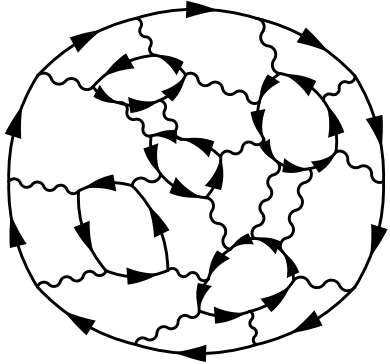} 
\caption{A typical vacuum planar diagram which is of the order of $N^0$.
}
\label{fig:vacuum_planar}
\end{figure}

\begin{figure}
\centering
        \includegraphics[height=5.0cm,width=5.0cm,angle=-30]{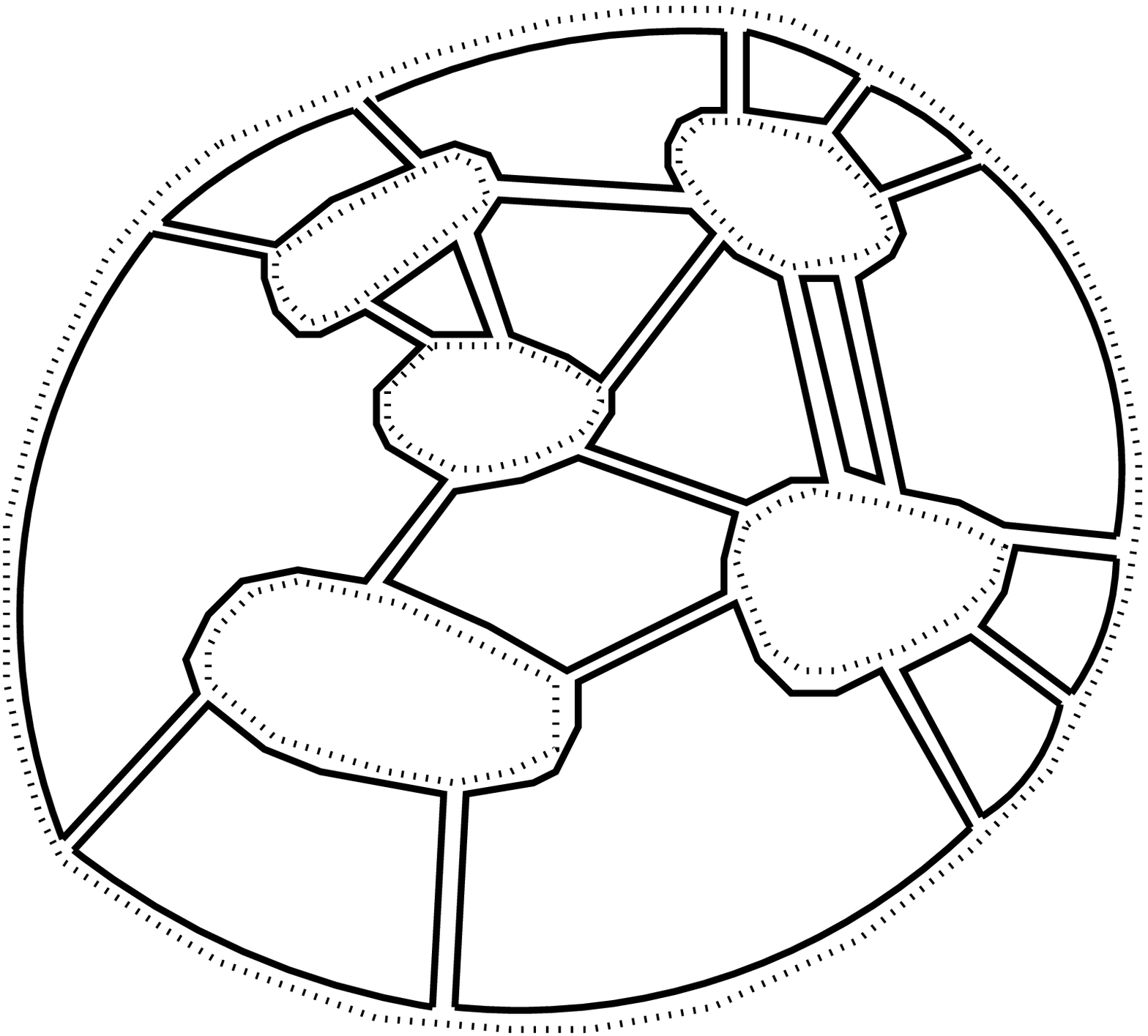} 
\caption{The full double line representation of 
the planar diagram shown in Fig. \ref{fig:vacuum_planar}.
The solid double lines represent the boson propagator
and double lines made of one solid and one dotted lines 
represent fermion propagators.
Loops of dotted lines are added to each fermion loops.
In this representation, there is a factor of $N$
for each closed single line loop 
whether it is a loop made of a solid or dotted line.
}
\label{fig:planar_double_double}
\end{figure}

The net power in $N$ for general Feynman diagrams
becomes 
\bqa
N^{-V/2 + L_f + \left[n + \frac{E_f + 2 E_b}{2} - 2 \right]},
\eqa
where $n$ is the number of single line loop,
$E_f$ ($E_b$) is the number of external 
lines for fermion (boson),
and $L_f$ is the number of fermion loop.
Here $[x]=x$ for non-negative $x$
and $[x]=0$ for negative $x$.
For vacuum diagram, the $N$-counting 
depends only on the topology 
of the Feynman diagram and it becomes
\bqa
N^{-2 g},
\label{power1}
\eqa
where $g$ is the genus of the 2d surface
on which Feynman graph is drawn using the full 
double line representation without any crossing.
In the full double line representation,
we draw not only the boson propagator as a double
line, but also the fermion propagator as a double
line where the additional line associated with
a fermion loop corresponds to the flavor degree
of freedom which run from $1$ to $N$.
As a result, all planar diagrams are generically order of $N^0$.
Some typical planar diagrams are shown in Figs. \ref{fig:vacuum_planar} and \ref{fig:planar_double_double}.

The above counting is based on the local
consideration on the space of internal momenta
near the singular manifold.
It turns out that planar diagrams precisely
follow the counting obtained from this local considerations.
Some non-planar diagrams may deviate from this local counting
by some factor of $\log N$.
The full structure of non-planar diagrams are not completely
understood yet.

Power counting of diagrams with external lines
can be easily obtained from the counting of vacuum diagrams.
The leading contributions come from the 
planar diagrams where the genus of the underlying
2d surface is zero.
In principle, there can be infinitely many diagrams
which are order of $N^0$ ($N^{-1}$) for the boson 
(fermion) self energy and $N^{-1/2}$ 
for the three-point vertex function.

Because the one-patch theory 
is a chiral theory,
there are strong kinematic constraints.
Using this, one can prove 
that all planar diagrams for boson self energy 
 vanish beyond the one loop.
Moreover one can show that the beta function 
vanishes and fermions have no anomalous scaling dimension
to the leading order of $N$ (the contributions of planar diagrams).
However, there are still infinitely many non-vanishing
planar diagrams for the fermion self energy and the vertex function.

Although the present theory has fermions with vector flavors,
it behaves like a matrix model in the large $N$ limit.
The reason why the genus expansion arises from this vector model
is that the angle around the Fermi surface plays
the role of an additional flavor.
In the usual matrix model, $N$ controls the genus expansion
and the t'Hooft coupling $\lambda$ controls the loop expansion.
One expects that a continuum world sheet description of 
string emerges in a matrix model
when both $N$ and  $\lambda$
are large.
In the present case,
the effective 't Hooft coupling is
order of $1$.
In other words, two diagrams with $L$-loops
and $(L+1)$-loops have the same order of magnitude
as far as they have same topology.
This is because there is no dimensionless parameter in the theory
other than $N$ as discussed earlier.
With $\lambda \sim 1$, it is unlikely that one can use 
a dual gravity description in a weakly curved space-time
to understand this non-Fermi liquid state.
Recently, it has been pointed out that
the non-Fermi liquid states studied from classical gravity
\cite{REY,H1,H2,H3,H4,H5,H6,H7,H8,H9}
show phenomenology of a system where
Fermi surface is coupled with a large number of localized
bosonic degrees of freedom\cite{H10}.

One may view the present theory as a weakly coupled string theory
in a highly curved background.
It would be interesting to find a way to stabilize the theory
at a large 't Hooft coupling.
However, this appears to be difficult to achieve
in non-supersymmetric theories due to instabilities.

If one includes two patches of Fermi surface,
even non-planar diagrams are important due to 
UV divergence, and the genus expansion appears to break down\cite{MAX}.
It is yet to be understood how Feynman diagrams 
are organized in this case. 
On the other hand, one can perturbatively control the theory
if one introduces an additional control parameter besides $1/N$\cite{Mross}.

\section{Holographic description of quantum field theory}

Although the AdS/CFT correspondence has been conjectured
based on the superstring theory,
it is possible that the underlying holographic principle 
is more general
and a wider class of 
quantum field theories
can be understood 
through holographic descriptions\cite{KLEBANOV,DAS,Gopakumar:2004qb,POLCHINSKI09,HOLO}.
According to the holographic principle,
the partition function of a D-dimensional theory
can be written as a partition function of a (D+1)-dimensional theory
where the information on the D-dimensional theory is encoded
through boundary conditions as
\bqa
Z[J(x)] & = & \int D \phi(x) e^{-S^{D}[\phi]-\int dx J \phi } \nn
& = & \left. \int D ~``J(x,z)" e^{-S^{(D+1)}[J(x,z)] } \right|_{J(x,0) = J(x)}, \nn
\eqa
where $J(x,z)$ represent (D+1)-dimensional degrees of freedom. 
It is most likely that the holographic description 
will be useful only for quantum field theories 
which satisfy certain conditions, such as 
factorization of correlation functions and 
a large gap in the spectrum of scaling dimensions.
Nevertheless, it would be still useful to develop
a general prescription for the mapping.
In this lecture, 
we will present a prescription
that may be useful in deriving holographic theory
for general quantum field theory\cite{HOLO}.

\subsection{Toy-model : $0$-dimensional scalar theory}

To illustrate the basic idea, 
we first consider the simplest field theory : 
$0$-dimensional scalar theory.
In zero dimension, the partition function
is given by an ordinary integration, 
\bqa
Z[\JJ] & = & \int d \Phi ~e^{-S [\Phi]}.
\label{ZJ}
\eqa
We consider an action $S[\Phi] = S_M[\Phi] + S_\JJ[\Phi]$ with
\bqa
S_{M}[\Phi] &=& M^2 \Phi^2, \nn
S_\JJ[\Phi] &=& \sum_{n=1}^\infty \JJ_n \Phi^n.
\eqa
Here $S_M$ is the bare action with `mass' $M$.
$S_\JJ$ is a deformation with sources $\JJ_n$.
For simplicity,
we will consider deformations
upto quartic order : $\JJ_n = 0$ for $n>4$.
Here is the prescription to construct a holographic theory.

\begin{itemize}
\item Step 1. Introduce an auxiliary field\cite{POLONYI}.

We add an auxiliary field $\td \Phi$ with mass $\mu$,
\bqa
Z[\JJ] & = & \mu \int d \Phi d \tilde \Phi ~e^{-(S[\Phi] + \mu^2 \tilde \Phi^2)}.
\label{eq:ax}
\eqa
Then, we find a new basis $\phi$ and $\tilde \phi$ 
\bqa
\Phi &=& \phi + \tilde \phi, \nn
\tilde \Phi &=& A \phi + B \tilde \phi,
\eqa
in such a way that
the `low energy field' $\phi$ has a mass $M^{'}$
which is slightly larger than the original mass $M$,
and the `high energy field' $\tilde \phi$ has a large mass $m^{'}$,
\bqa
M^{'2} &=& M^2 e^{2 \alpha dz}, \nn
m^{'2} &=& \frac{M^2}{2\alpha dz},
\eqa
where $dz$ is an infinitesimally small parameter
and $\alpha$ is a positive constant.
Quantum fluctuations for $\phi$ become
slightly smaller than the original field $\Phi$.
The missing quantum fluctuations are 
compensated by the high energy field $\tilde \phi$.
In terms of the new variables, 
the partition function is written as
\bqa
Z[\JJ] & = &  \left( \frac{M m^{'}}{ M^{'}}  + \frac{M M^{'}}{ m^{'} }   \right) 
\int d \phi d \tilde \phi ~e^{-(S_\JJ[\phi + \tilde \phi] + M^{'2} \phi^2 + m^{'2} \tilde \phi^2)}.
\eqa

\item Step 2. Rescale the fields.

To maintain the same form for the quadratic action,
we rescale the fields as
\bqa
\phi  \rightarrow  e^{- \alpha dz } \phi, ~~~~ 
\td \phi  \rightarrow  e^{- \alpha dz } \td \phi.
\eqa
Then the couplings are rescaled as
$\JJ_n \rightarrow j_n = \JJ_n e^{- n \alpha dz }$ 
and $m \rightarrow  m^{'} e^{-\alpha dz}$.

\item Step 3. Expand the action in the power series of the low energy field.

The new action becomes
\bqa
S_j[\phi + \td \phi] &   = & S_j[\td \phi]  
 + ( j_1 + 2 j_2 \td \phi + 3 j_3 \td \phi^2 + 4 j_4 \td \phi^3 ) \phi \nn
&& + ( j_2 + 3 j_3 \td \phi + 6 j_4 \td \phi^2 ) \phi^2 
 + ( j_3  + 4 j_4 \td \phi ) \phi^3 
 + j_4 \phi^4. 
\label{expansion}
\eqa
In the standard renormalization group (RG) procedure\cite{POLCHINSKI84},
one integrates out the high energy field 
to obtain an effective action for the low energy field
with renormalized coupling constants.
Here we take an alternative view 
and interpret the high energy field $\td \phi$ 
as fluctuating sources for the low energy field.
This means that the sources for the low energy field 
can be regarded as dynamical fields 
instead of fixed coupling constants. 

\item Step 4. Decouple low energy field and high energy field.

We decouple the high energy field and the low energy field
by introducing Hubbard-Stratonovich fields $J_n$ and $P_n$,
\bqa
Z[\JJ] & = &   m  
\int d \phi d \tilde \phi  \Pi_{n=1}^4 (d J_n d P_n ) ~ 
e^{-(S_j^{'} + M^2 \phi^2 + m^2 \tilde \phi^2)},
\eqa
where
\bqa
S_j^{'} & = &  S_j[ \td \phi]  \nn
&& + i P_1 J_1 - i P_1 ( j_1 + 2 j_2 \td \phi + 3 j_3 \td \phi^2 + 4 j_4 \td \phi^3 ) + J_1 \phi \nn
&& + i P_2 J_2 - i P_2 ( j_2 + 3 j_3 \td \phi + 6 j_4 \td \phi^2 ) + J_2 \phi^2 \nn
&& + i P_3 J_3 - i P_3 ( j_3  + 4 j_4 \td \phi ) + J_3 \phi^3 \nn
&& + i P_4 J_4 - i P_4 j_4 + J_4 \phi^4.
\label{Si}
\eqa

\item Step 5. Integrate out the high energy mode.

We integrate out $\td \phi$ to the order of $dz$.
The auxiliary fields $P$ and $J$
acquire non-trivial action,
\bqa
Z[\JJ] & = &  
\int d \phi \Pi_{n=1}^4 (d J_n d P_n ) ~ e^{-(S_{J}[\phi] + M^2 \phi^2 + S^{(1)}[J,P])},
\label{eq:dz}
\eqa
where
\bqa
S^{(1)}[J,P] 
&= &  \sum_{n=1}^4 i ( J_n - \JJ_n + n \alpha dz \JJ_n ) P_n  \nn
&& + \frac{\alpha dz}{2 M^2} ( i \td \JJ_1 + 2  P_1 \td \JJ_2 + 3  P_2 \td \JJ_3 + 4  P_3 \td \JJ_4 )^2,
\label{S1}
\eqa
and $\td \JJ_n = (\JJ_n + J_n)/2$.
One can explicitly check that the above action
reproduces all renormalized coupling to the order of $dz$ 
if $P_n$ and $J_n$ are integrated out.

\item Step 6. Repeat the steps 1-5 for the low energy field

If we keep applying the same procedure to the low energy field,
the partition function can be written as a functional integration
over fluctuating sources and conjugate fields,
\bqa
Z[\JJ] & = &    
\int \Pi_{n=1}^4 (D J_n D P_n ) ~ e^{-S[J,P]},
\label{Zgravity}
\eqa
where
\bqa
S[J,P] 
&= & \int_0^\infty dz \Bigl[  i ( \partial_z J_n + n \alpha J_n ) P_n  \nn
&& + \frac{\alpha}{2M^2} ( i J_1 + 2  P_1 J_2 + 3  P_2 J_3 + 4  P_3 J_4 )^2  \Bigr]. 
\label{Sgravity}
\eqa
Here $DJ_n DP_n$ represent functional integrations 
over one dimensional fields $J_n(z), P_n(z)$ 
which are defined on the semi-infinite line $[0,\infty)$.
The boundary value of $J_n(z)$ is 
fixed by the coupling constants of 
the original theory, $J_n(0) = \JJ_n$.
$P_n(z)$ is the conjugate field of $J_n(z)$
which describes
physical fluctuations of the operator $\phi^n$.
This can be seen from the equations of motion for $J_n$.

\end{itemize}

The theory given by Eqs. (\ref{Zgravity}) and (\ref{Sgravity}),
which is exactly dual to the original theory,
is an one-dimensional local quantum theory.
The emergent dimension $z$ corresponds to logarithmic energy scale\cite{VERLINDE}.
The parameter $\alpha$ determines the rate the energy scale is changed.
The partition function in Eq. (\ref{Zgravity})
does not depend on the rate high energy modes are eliminated
as far as all modes are eventually eliminated.
Moreover, at each step of mode elimination, 
one could have chosen $\alpha$ differently.
Therefore, $\alpha$ can be regarded as a function of $z$.
If one interprets $z$  as `time', 
it is natural to identify $\alpha(z)$ 
as the `lapse function',
that is, $\alpha(z) = \sqrt{ g_{zz}(z) }$,
where $g_{zz}(z)$ is the metric.
Then one can view Eqs. (\ref{Zgravity}) and (\ref{Sgravity}) as
an one-dimensional gravitational theory with matter fields $J_n$.
This becomes more clear 
if we write the Lagrangian as
\bqa
L = P_n \partial_z J_n - \alpha H,
\eqa
where $H$ is the Hamiltonian
(the reason why $H$ is not Hermitian  
is that we started from the Euclidean field theory).
However, there is one difference from 
the usual gravitational theory.
In the Hamiltonian formalism of gravity\cite{ADM},
the lapse function is a Lagrangian multiplier
which imposes the constraint $H=0$.
However, in Eq. (\ref{Zgravity}), $\alpha$ is not integrated over
and the Hamiltonian constraint is not imposed.
This is due to the presence of the boundary at $z=0$
which explicitly breaks the reparametrization symmetry.
In particular, the `proper time' 
from $z=0$  to $z=\infty$ given by
\bq 
l = \int_0^\infty \alpha(z) dz
\eq
is a quantity of physical significance
which measures the total warping factor.
To reproduce the original partition function in Eq. (\ref{ZJ})
from Eq. (\ref{Zgravity}),
one has to make sure that $l=\infty$ to include
all modes in the infrared limit.
Therefore, $l$ should be fixed to be infinite.
As a result, one should not integrate over all possible
$\alpha(z)$ some of which give different $l$.
This is the physical reason 
why the Hamiltonian constraint is not imposed
in the present theory.
This theory can be viewed as a gravitational theory
with the fixed size along the $z$ direction.

Although there are many fields in the bulk, i.e. $J_n, P_n$ for each $n$, 
there is only one propagating mode,
and the remaining fields are non-dynamical in the sense they 
strictly obey constraints imposed by their conjugate fields.
This is not surprising because we started with one dynamical field $\Phi$.
There is a freedom in choosing one independent field.
In this case, it is convenient to choose $J_3$ 
as an independent field.
If one eliminates all dependent fields,
one can obtain the local bulk action for one independent field.

\subsection{$D$-dimensional O(N) vector theory}

The same procedure can be generalized to 
$D$-dimensional field theory.
For example, the partition function for
the $D$-dimensional O(N) vector model,
\bqa
S[\Phi] &=& \int d {\bf x} d {\bf y} ~\Phi_a({\bf x}) G_M^{-1}({\bf x}-{\bf y}) \Phi_a({\bf y}) \nn
 && + \int d {\bf x} ~ \Bigl[
\JJ_a \Phi_a + \JJ_{ab} \Phi_a \Phi_b + \JJ_{abc} \Phi_a \Phi_b \Phi_c + \JJ_{abcd} \Phi_a \Phi_b \Phi_c \Phi_d \nn
&& + \JJ_{ab}^{ij}  \partial_i \Phi_a \partial_j \Phi_b
+ \JJ_{abc}^{ij} \Phi_a \partial_i \Phi_b \partial_j \Phi_c
+ \JJ_{abcd}^{ij} \Phi_a \Phi_b \partial_i \Phi_c \partial_j \Phi_d
\Bigr]
\label{ONS}
\eqa
can be written as a $(D+1)$-dimensional functional integration,
\bqa
Z[\JJ] & = &    
\int D J D P ~ e^{-S[J,P]},
\label{Zgravity5}
\eqa
where the bulk action is given by
\bqa
S[J,P] & =  \int d {\bf x} dz & 
\Bigl\{  
i P_a ( \partial J_a - \frac{2+D}{2} \alpha J_a )
+i P_{ab} ( \partial J_{ab} - 2 \alpha J_{ab} )
+i P_{ab,ij} ( \partial J_{ab}^{ij} ) \nn
&& + i P_{abc} ( \partial J_{abc} - \frac{6-D}{2} \alpha J_{abc} )
+i P_{abc,ij} ( \partial J_{abc}^{ij} - \frac{2-d}{2} \alpha J_{abc}^{ij} ) \nn
&& 
+i P_{abcd} ( \partial J_{abcd} - (4-D) \alpha J_{abcd} )
+i P_{abcd,ij} ( \partial J_{abcd}^{ij} - (2-d) \alpha J_{abcd}^{ij} )
\Bigr\} \nn
& + \frac{1}{4} \int d {\bf x} d {\bf y} dz & 
\Bigl\{  
\alpha s_a({\bf x}) G^{'}({\bf x}-{\bf y}) s_a({\bf y})
\Bigr\},
\label{Sgravity5}
\eqa
with
\bqa
s_a & = &
\Bigl[
 i J_a 
+ 2  P_b J_{ab} 
- 2  \partial_j ( J_{ab}^{ij} \partial_i P_b )
+ 3  P_{bc} J_{abc} 
-  \partial_j ( J_{abc}^{ij} \partial_i P_{bc} ) \nn
&& ~~ 
+  P_{bc,ij} J_{abc}^{ij}
+ 4  P_{bcd} J_{abcd} 
- \frac{2}{3}  \partial_j ( J_{abcd}^{ij} \partial_i P_{bcd} )
+ 2  J_{abcd}^{ij} P_{bcd,ij}
\Bigr],
\eqa 
$G^{'}({\bf x}) \equiv  M \partial_M G_M({\bf x})$, 
and
$ \partial = \frac{\partial}{\partial z} - 
\alpha \sum_{i=1}^D  x_i \frac{\partial}{\partial x_i}$.
Here 
$\int d {\bf x}$ 
and $\int d {\bf y}$ 
are integrations 
on a $D$-dimensional manifold ${\cal M}^D$, and
$G_M^{-1}({\bf x})$ is the regularized kinetic energy with cut-off $M$.

The dual theory
is given by the 
functional integrals of the source fields $J$ and their
conjugate fields $P$ in the $(D+1)$-dimensional space
${\cal M}^D \times [0,\infty)$
with the boundary condition $J({\bf x},z=0) = \JJ({\bf x})$. 
If the $D$-dimensional manifold ${\cal M}^D$ has a finite volume $V$,
the volume at scale $z$ is given by $V e^{-\alpha D z}$.

One key difference from the $0$-dimensional theory is that
there exist bulk fields with non-trivial spins.
In Eq. (\ref{Sgravity5}), there are spin two fields
which are coupled to the energy momentum tensor at the boundary.
In the presence of more general deformations in the boundary theory,
one needs to introduce fields with higher spins\cite{VASILIEV,YIN}.

One can decompose the tensor sources into singlets and 
traceless parts
and take the large $N$ limit
where saddle point solutions
become exact 
for singlet fields.
One can integrate out all non-singlet fields
and obtain an effective theory for single fields alone.
However, the resulting effective action for single fields
become non-local in this $O(N)$ vector model.
This is because there are light non-singlet fields
in the bulk and integrating over those soft modes
generates non-local correlations for singlet fields.
This means that we should keep light non-singlet fields
as `low energy degrees of freedom' in the bulk description 
if we want to use a local description.

\subsection{Phase Transition and Critical Behaviors }

\begin{figure}
\centering
        \includegraphics[height=7cm,width=15cm]{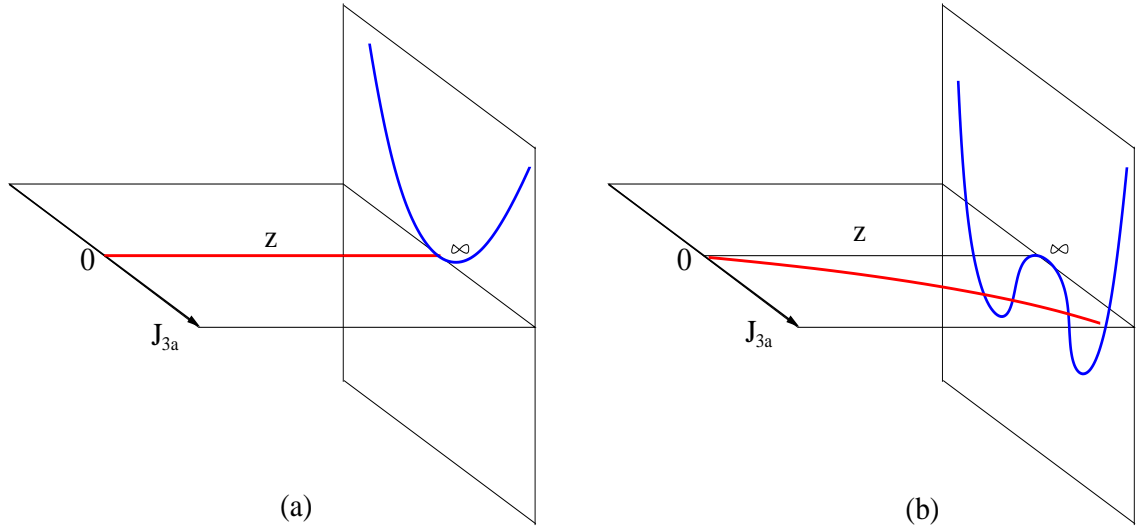} 
\caption{
Saddle point configuration for a non-singlet source field $J_{3a}(z)$ 
(a) in the disordered phase 
and (b) in the ordered phase.
When $\JJ_2$ is sufficiently negative,
a Mexican-hat potential at the IR boundary 
drags $J_{3a}(z)$ away from $J_{3a}(z)=0$ in the bulk.
At the critical point, $J_{3a}$ at the IR boundary $z=\infty$
is more or less free to fluctuate, 
generating algebraic correlations between 
fields inserted at the UV boundary $z=0$.
}
\label{fig:SSB}
\end{figure}

One can understand 
the phase transition
and the critical properties of the model in $D>2$
using the holographic theory.
As one tunes the singlet sources at the UV boundary,
the shape of potential in the IR limit changes accordingly.
In some parameter regime,
the bulk fields are forced to spontaneously break
the $O(N)$ symmetry.
This is illustrated in Fig. \ref{fig:SSB}.

One can also compute correlation functions of 
the singlet operators at the critical point
using the similar method used in the AdS/CFT correspondence.
For this one integrates over all bulk fields
consistent with the $x$-dependent UV boundary condition. 
The bulk action can be computed as a function 
of the UV sources and this gives
the generating function for the boundary theory.
From this one can compute the critical exponents
of singlet operators, which matches with the known field
theory predictions.
More recently, the present prescription has been applied
to large $N$ gauge theory 
where a field theory of closed loops
arise as a holographic dual for 
the $U(N)$ gauge theory\cite{LEE2010}.

\section{Acknowledgment}
This note is based on the lectures given at TASI in June 2010.
I would like to thank Thomas Banks, Michael Dine and Subir Sachdev 
for their kind invitation to give lectures at TASI.
I am also grateful to Thomas DeGrand, K.T. Mahanthappa and Susan Spika
for the hospitality during the summer school.
Parts of this material were also presented at the 
APCTP Focus Program on Aspects of Holography and Gauge/string duality
held at Pohang, Korea in August 2010.
I thank Deog Ki Hong, Sang-Jin Sin and Piljin Yi
for giving me the opportunity to give lectures.
I would like to thank lecturers, participants and students 
of TASI and APCTP workshop for stimulating comments and discussions.
Finally, I thank
Andrey Chubukov,
Guido Festuccia,
Matthew Fisher,
Sean Hartnoll,
Michael Hermele,
Yong Baek Kim,
Patrick Lee,
Hong Liu,
Max Metlitski,
Lesik Motrunich,
Joe Polchinski,
Subir Sachdev,
T. Senthil,
Mithat Unsal
and
Xiao-Gang Wen
for many illuminating discussions
in the past.
This work was supported by NSERC.

\end{document}